\documentclass{article}
\usepackage{url}
\usepackage[utf8]{inputenc}
\usepackage{ismir,amsmath,cite}
\usepackage[bookmarks=false]{hyperref}
\usepackage{graphicx}
\usepackage{color}

\usepackage{tikz}
\newcommand{\ExternalLink}{%
    \tikz[x=1.2ex, y=1.2ex, baseline=-0.05ex]{%
        \begin{scope}[x=1ex, y=1ex]
            \clip (-0.1,-0.1) 
                --++ (-0, 1.2) 
                --++ (0.6, 0) 
                --++ (0, -0.6) 
                --++ (0.6, 0) 
                --++ (0, -1);
            \path[draw, 
                line width = 0.5, 
                rounded corners=0.5] 
                (0,0) rectangle (1,1);
        \end{scope}
        \path[draw, line width = 0.5] (0.5, 0.5) 
            -- (1, 1);
        \path[draw, line width = 0.5] (0.6, 1) 
            -- (1, 1) -- (1, 0.6);
        }
    }

\title{Towards Music Captioning:\\Generating Music Playlist Descriptions}
\twoauthors
  {Keunwoo Choi, Gy\"orgy Fazekas, Mark Sandler} {Centre for Digital Music\\Queen Mary University of London\\{\tt keunwoo.choi\tt @qmul.ac.uk}}
  {Brian McFee, Kyunghyun Cho} {Center for Data Science\\New York University\\{\tt \{first.last\}\tt @nyu.edu}}

\renewcommand{\vec}[1]{\mathbf{#1}}
\begin{document}

\maketitle
\begin{abstract}
Descriptions are often provided along with recommendations to help users' discovery. Recommending automatically generated music playlists (e.g. personalised playlists) introduces the problem of generating descriptions. In this paper, we propose a method for generating music playlist descriptions, which is called as music captioning. In the proposed method, audio content analysis and natural language processing are adopted to utilise the information of each track. 
\end{abstract}
\section{Introduction}
\textbf{Motivation:} 
One of the crucial problems in music discovery is to deliver the summary of music without playing it. One common method is to add descriptions of a music item or playlist, e.g. 
\textit{Getting emotional with the undisputed King of Pop}\footnote{\href{https://itunes.apple.com/us/playlist/michael-jackson-love-songs/idpl.8058d87c60b647a7bc81185b9f59e4c2}{\textit{Michael Jackson: Love songs and ballads} by \textit{Apple Music} \ExternalLink}}, 
\textit{Just the right blend of chilled-out acoustic songs to work, relax, think, and dream to\footnote{\href{https://open.spotify.com/user/spotify_uk_/playlist/48910w3L1DNiqvMHbUfZyY}{\textit{Your Coffee Break} by \textit{Spotify} \ExternalLink} }}.
 These examples show that they are more than simple descriptions and even add value to the curated playlist as a product.
\begin{figure}[t!]
\begin{center} \includegraphics{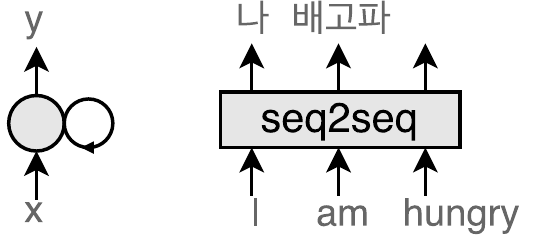} \end{center}
 \caption{A block diagram of an RNN unit (left) and sequence-to-sequence module that is applied to English-Korean translation (right).}
 \label{fig:seq2seq}
\end{figure}

There have been attempts to automate the generation of these descriptions. In \cite{eck2008automatic}, Eck et al. proposed to use social tags to describe each music item. Fields proposed a similar idea for playlist using social tag and topic model \cite{fields2010using} using Latent Dirichlet Allocation \cite{blei2003latent}. Besides text, Bogdanov introduced music avatars, whose outlook - hair style, clothes, and accessories - describes the recommended music \cite{bogdanov2013semantic}.

\textbf{Background:}
$\bullet$ \textit{RNNs}: RNNs are neural networks that have a unit with a recurrent connection, whose output is connect to the input of the unit (Figure \ref{fig:seq2seq}, left). They currently show state-of-the-art performances in tasks that involve sequence modelling. Two types of RNN unit are widely used: Long Short-Term Memory (LSTM) unit \cite{hochreiter1997long} and Gated Recurrent Unit (GRU)\cite{cho2014properties}. 

$\bullet$ \textit{Seq2seq}: Sequence-to-sequence (seq2seq) learning indicates training a model whose input and output are sequences (Figure \ref{fig:seq2seq}, right). Seq2seq models can be used to machine translation, where a phrase in a language is summarised by an encoder RNN, which is followed by a decoder RNN to generate a phrase in another language \cite{cho2014learning}.

$\bullet$ \textit{Word2vec}: Word embeddings are distributed vector representations of words that aim to preserve the semantic relationships among words. One successful example is \textit{word2vec} algorithm, which is usually trained with large corpora in an unsupervised manner \cite{mikolov2013distributed}.

$\bullet$ \textit{ConvNets}: Convolutional neural networks (ConvNets) have been extensively adopted in nearly every computer vision task and algorithm since the record-breaking performance of AlexNet \cite{krizhevsky2012imagenet}. ConvNets also show state-of-the-art results in many music information retrieval tasks including auto-tagging \cite{choi2016automatic}.

\section{Problem Definition}
The problem of \textit{music captioning} can be defined as \textit{generating a description for a set of music items using on their audio content and text data.} When the set includes more than one item, it can be also called as \textit{music playlist captioning}.

\begin{figure}[t!]
\begin{center} \includegraphics{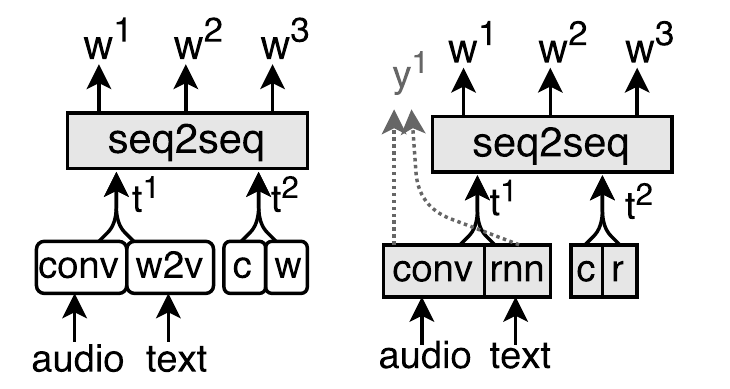} \end{center}
 \caption{The diagrams of two proposed approaches, where coloured blocks indicate trainable modules. The first approach uses a pre-trained ConvNet (\textit{conv}) and \textit{word2vec} (\textit{w2v}) and only sequence-to-sequence model is trained. In the second approach, the whole blocks are trained - a ConvNet to summarise the audio content, an RNN to summarise the text data of each track. An additional labels (\textit{y}) such as genres or tags can be provided to help the training.}
 \label{fig:example}
\end{figure}

\section{The proposed method}
Both of the approaches use sequence-to-sequence model, as illustrated in Figure \ref{fig:example}. In the sequence-to-sequence model, the encoder consists of two-layer RNN with GRU and encodes the track features into a vector, i.e., the encoded vector summarises the information of the input. This vector is also called \textit{context vector} because it provides context information to the decoder. The decoder consists of two-layer RNN with GRU and decodes the context vector to a sequence of word or word embeddings. The models are written in Keras and uploaded online\footnote{\url{http://github.com/keunwoochoi/ismir2016-ldb-audio-captioning-model-keras}}\cite{chollet2015keras}.

\subsection{Pre-training approach}
This approach takes advantage of a pre-trained word embedding model\footnote{\url{https://radimrehurek.com/gensim/models/word2vec.html}} and a pre-trained auto-tagger\footnote{\url{https://github.com/keunwoochoi/music-auto_tagging-keras}, \cite{choi2016automatic}}. Therefore, the number of parameters to learn is reduced while leveraging additional data to train word-embedding and auto-tagger. Each data sample consists of a sequence of $N$ track features as input and an output word sequence length of $M$, which is an album feature. 

\textbf{Input/Output\footnote{The dimensions can vary, we describe in details for better understanding.}:} A $n$-th track feature, $\vec{t^n} \in \mathbf{R}^{350}$, represents one track and is created by concatenating the audio feature, $\vec{t_a^n}\in \mathbf{R}^{50}$, and the word feature, $\vec{t_w^n}\in \mathbf{R}^{300}$, i.e. $\vec{t}$ = {[}$\vec{t_a}$;$\vec{t_w}${]}. For computing $\vec{t_a}$, a convolutional neural network that is trained to predict tags is used to output 50-dim vector for each track \cite{choi2016automatic}. $\vec{t_w}$ is computed by $\sum_k {\vec{w_k}}/K$, where $\vec{w_k}$ refers to the embedding of $k$-th word in the metadata\footnote{ Because these word embeddings are distributed representations in a semantic vector space, average of the words can summarise a bag of words and was used as a baseline in sentence and paragraph representation \cite{dai2015document}.}. The word embedding were trained by \textit{word2vec} algorithms and Google news dataset \cite{mikolov2013distributed}.

An playlist feature is a sequence of word embeddings of the playlist description, i.e. $\vec{p}=[{\vec{w^m}}]_{m=0,1,..m-1}$.
\subsection{Fully-training approach}
The model in this approach includes the training of a ConvNet for audio summarisation and an RNN for text summarisation of each track. The structure of ConvNet can be similar to the pre-trained one. The RNN module is trained to summarise the text of each track and outputs a \textit{sentence vector}. These networks can be provided with additional labels (notated as $y$ in the figure \ref{fig:example}) to help the training, e.g., genres or tags. In that case, the objective of the whole structure consists of two different tasks and therefore the training can be more regulated and stable.

Since the audio and text summarisation modules are trainable, they can be more relevant to the captioning task. However, this flexibility requires more training data.

\section{Experiments and Conclusions}
We tested the pre-training approach with a private production music dataset. The dataset has 374 albums and 17,354 tracks with descriptions of tracks, albums, audio signal and metadata. The learning rate is controlled by ADAM \cite{kingma2014adam} with an objective function of \textit{1-cosine proximity}. The model was trained to predict the album descriptions.

The model currently overfits and fails to generate correct sentences. One example of generated word sequence is \textit{dramatic motivating the intense epic action adventure soaring soaring soaring gloriously Roger\_Deakins\_cinematography Maryse\_Alberti}. This is expected since there are only 374 output sequences in the dataset -- if we use early stopping, the model underfits, otherwise it overfits. 

In the future, we plan to solve the current problem -- lack of data. The sentence generation can be partly trained by (music) corpora. A word2vec model that is trained with music corpora can be used to reduce the embedding dimension \cite{oramas2016natural}. The model can also be modified in the sense that the audio feature is optional and it mainly relies on metadata. In that case, acquisition of training data becomes more feasible. 
\section{Acknowledgements}
This work was part funded by the 
FAST IMPACt EPSRC Grant EP/L019981/1 and the European Commission H2020 research and innovation grant AudioCommons (688382). Mark Sandler acknowledges the support of the Royal Society as a recipient of a Wolfson Research Merit Award.
Brian McFee is supported by the Moore Sloan Data Science Environment at NYU. Kyunghyun Cho thanks the support by Facebook, Google (Google Faculty Award 2016) and NVidia (GPU Center of Excellence 2015-2016). The work is done during Keunwoo Choi is visiting Center for Data Science in New York University. 

\bibliography{audio_captioning}

\begin{thebibliography}{10}

\bibitem{blei2003latent}
David~M Blei, Andrew~Y Ng, and Michael~I Jordan.
\newblock Latent dirichlet allocation.
\newblock {\em Journal of machine Learning research}, 3(Jan):993--1022, 2003.

\bibitem{bogdanov2013semantic}
Dmitry Bogdanov, Mart{\'\i}N Haro, Ferdinand Fuhrmann, Anna Xamb{\'o}, Emilia
  G{\'o}mez, and Perfecto Herrera.
\newblock Semantic audio content-based music recommendation and visualization
  based on user preference examples.
\newblock {\em Information Processing \& Management}, 49(1):13--33, 2013.

\bibitem{cho2014properties}
Kyunghyun Cho, Bart Van~Merri{\"e}nboer, Dzmitry Bahdanau, and Yoshua Bengio.
\newblock On the properties of neural machine translation: Encoder-decoder
  approaches.
\newblock {\em arXiv preprint arXiv:1409.1259}, 2014.

\bibitem{cho2014learning}
Kyunghyun Cho, Bart Van~Merri{\"e}nboer, Caglar Gulcehre, Dzmitry Bahdanau,
  Fethi Bougares, Holger Schwenk, and Yoshua Bengio.
\newblock Learning phrase representations using rnn encoder-decoder for
  statistical machine translation.
\newblock {\em arXiv preprint arXiv:1406.1078}, 2014.

\bibitem{choi2016automatic}
Keunwoo Choi, George Fazekas, and Mark Sandler.
\newblock Automatic tagging using deep convolutional neural networks.
\newblock In {\em International Society of Music Information Retrieval
  Conference}. ISMIR, 2016.

\bibitem{chollet2015keras}
Fran{\c{c}}ois Chollet.
\newblock Keras.
\newblock {\em GitHub repository: https://github. com/fchollet/keras}, 2015.

\bibitem{dai2015document}
Andrew~M Dai, Christopher Olah, and Quoc~V Le.
\newblock Document embedding with paragraph vectors.
\newblock {\em arXiv preprint arXiv:1507.07998}, 2015.

\bibitem{eck2008automatic}
Douglas Eck, Paul Lamere, Thierry Bertin-Mahieux, and Stephen Green.
\newblock Automatic generation of social tags for music recommendation.
\newblock In {\em Advances in neural information processing systems}, pages
  385--392, 2008.

\bibitem{fields2010using}
Ben Fields, Christophe Rhodes, Mark d'Inverno, et~al.
\newblock Using song social tags and topic models to describe and compare
  playlists.
\newblock In {\em 1st Workshop On Music Recommendation And Discovery (WOMRAD),
  ACM RecSys, 2010, Barcelona, Spain}, 2010.

\bibitem{hochreiter1997long}
Sepp Hochreiter and J{\"u}rgen Schmidhuber.
\newblock Long short-term memory.
\newblock {\em Neural computation}, 9(8):1735--1780, 1997.

\bibitem{kingma2014adam}
Diederik Kingma and Jimmy Ba.
\newblock Adam: A method for stochastic optimization.
\newblock {\em arXiv preprint arXiv:1412.6980}, 2014.

\bibitem{krizhevsky2012imagenet}
Alex Krizhevsky, Ilya Sutskever, and Geoffrey~E Hinton.
\newblock Imagenet classification with deep convolutional neural networks.
\newblock In {\em Advances in neural information processing systems}, pages
  1097--1105, 2012.

\bibitem{mikolov2013distributed}
T~Mikolov and J~Dean.
\newblock Distributed representations of words and phrases and their
  compositionality.
\newblock {\em Advances in neural information processing systems}, 2013.

\bibitem{oramas2016natural}
Sergio Oramas, Luies Espinosa-Anke, Shuo Zhang, Horacio Saggion, and Xavier
  Serra.
\newblock Natural language processing for music information retrieval.
\newblock In {\em 17th International Society for Music Information Retrieval
  Conference (ISMIR 2016)}, 2016.

\end{thebibliography}

\end{document}